\documentclass[twocolumn,aps,showpacs,prb]{revtex4-1}
\usepackage{graphicx}
\usepackage{epstopdf}
\usepackage{amsmath}
\usepackage{multirow}

\begin{document}

\title{The influence of spin-phonon coupling on antiferromagnetic spin fluctuations in FeSe under pressure: the First-principles calculations with van der Waals corrections}

\author{Qian-Qian Ye}
\author{Kai Liu}\email{kliu@ruc.edu.cn}
\author{Zhong-Yi Lu}\email{zlu@ruc.edu.cn}

\date{\today}

\affiliation{Department of Physics, Renmin University of China, Beijing 100872, China}

\begin{abstract}

The electronic structures, lattice dynamics, and magnetic properties of crystal $\beta$-FeSe under hydrostatic pressure have been studied by using the first-principles electronic structure calculations with van der Waals (vdW) corrections. With applied pressures, the energy bands around the Fermi energy level consisting mainly of Fe-$3d$ orbitals show obvious energy shifts and occupation variations, and meanwhile the frequencies of all optical phonon modes increase. Among these phonon modes, the $A_{1g}$ mode, which relates to the Se height from the Fe-Fe plane, shows a clear frequency jump in the range between 5 and 6 GPa. This is also the pressure range within which the highest superconducting transition temperature $T_c$ of FeSe is reached in experiments. In comparison with the other phonon modes, the zero-point vibration of the $A_{1g}$ mode induces the strongest fluctuation of local magnetic moment on Fe under a pressure from 0 to 9 GPa, and the induced fluctuation reaches the maximum around 5 GPa.  These results suggest that the effect of phonon via spin-phonon coupling need to be considered when exploring the superconducting mechanism in iron-based superconductors.
\end{abstract}



\maketitle

\section{INTRODUCTION}

With the simplest crystal structure among the iron-based superconductors ever found,\cite{sc1,sc2,sc3,sc4} PbO-type $\beta$-FeSe has attracted tremendous attention both experimentally and theoretically as an archetype system to explore the unconventional superconductivity mechanism in the iron-based superconductors. The superconducting transition temperature $T_c$ of FeSe is found to be 8 K at an ambient pressure,\cite{sc4} and extremely sensitive to external pressures while it can grow to the maximum of $\sim$37 K around 6$-$9 GPa.\cite{gpa1,gpa2,gpa3} On the other hand, the nuclear magnetic resonance (NMR) measurement has found strongly enhanced antiferromagnetic (AFM) spin fluctuations near $T_{c}$ and that both $T_{c}$ and spin fluctuations are raised by pressures, suggesting a close correlation between the spin fluctuations and the superconducting mechanism in the iron-based superconductors.\cite{nmr}

With applied pressures, the crystal structure of $\beta$-FeSe shows notable changes. \cite{gpa1,gpa2,gpa3} Its volume undergoes as high as 20\% reduction and the interlayer spacing shows large decreases.\cite{gpa1,gpa2} In contrast, the height of Se from the Fe-Fe plane first decreases and then increases with pressure.\cite{gpa2} By summarizing the $T_c$s of FeSe under pressures and the $T_c$s of various iron (and nickel)-based superconductors, a striking correlation between the $T_c$ and the height of anion (\textit{i.e.} Se in FeSe) is revealed.\cite{gpa3,mizuguchi} Moreover, Moon and Choi have found that the magnetic interactions and magnetic order are very sensitive to the height of chalcogen species from Fe-Fe plane by studying the FeTe system with fixed height $Z_{Te}$ using the density-functional calculations.\cite{moon} This is also reflected in the experiments that the nonsuperconducting FeTe bulk samples become superconducting in the FeTe thin films under tensile stress.\cite{cao} These observations indicate that the crystal structures and magnetic properties, both of which link with $T_{c}$ in iron chalcogenide superconductors, are obviously correlated and can be tuned by an applied pressure.

A direct consequence of substantial changes in the crystal structure under pressure is that the lattice dynamics would be influenced. However, within the standard McMillan-Eliashberg framework, the electron-phonon coupling calculations without spin-polarization give too low values for the transition temperature $T_{c}$ of FeSe.\cite{Subedi} When the spin polarization effects are included, the calculated electron-phonon coupling value in the checkerboard AFM N\'eel structure shows about a twofold increase, but still cannot account for the experimentally observed $T_{c}$ of FeSe.\cite{Bazhirov} It is now commonly accepted that FeSe is not a conventional electron-phonon superconductor,\cite{Subedi} but the spin-fluctuation mediated paring yields the unconventional superconductivity.\cite{rmp1,rmp2} Nevertheless, a clear iron isotope effect on $T_c$ of FeSe was observed in experimental measurement, \cite{Khasanov} and the magnetic properties were found to be very sensitive to the lattice parameters of FeSe$_x$Te$_{1-x}$ from the density functional calculations,\cite{moon} thus we are curious about what role the lattice vibrations play in the spin fluctuations (via the spin-phonon coupling effect) of FeSe. More importantly, how the role of the lattice vibrations will change along with pressure, especially around the pressure with the highest $T_c$ of FeSe, has never been studied.

As FeSe has Se-Fe$_{2}$-Se layers composed of edge-sharing tetrahedra with an Fe center, the van der Waals (vdW) interaction plays an important role in the interlayer bonding. When a pressure is applied, the interlayer distance of FeSe shows much larger reduction compared with the in-plane lattice constants.\cite{gpa1} In order to precisely describe the lattice dynamics of FeSe under pressure in calculations, the vdW interactions between FeSe layers need to be accurately accounted for. Since the conventional density functionals are unable to describe correctly the vdW interactions, which arise from nonlocal long-range electron correlations,\cite{Andersson,Langreth} both nonlocal correlation functional\cite{Dion,Soler,Klim,Lee} and a semi-empirical dispersion potential\cite{Wu,Grimme} method have been proposed to include the dispersion interactions. Nowadays, the density functional theory (DFT) problem for vdW interaction has become a very active field and theoretical studies have been carried out on various relevant molecules and materials.\cite{Grimme,Langreth,Klim2,Klim3,Bjorkman,Aradhya,hyldgaard,Berland} However, the DFT calculations with vdW-interaction corrections have rarely been applied to the iron-based superconductors.\cite{Ricci, Nakamura}

We have studied the electronic structures, lattice dynamics, and magnetic properties of FeSe under pressure using DFT calculations with vdW corrections. The variations of band structures and the phonon frequencies from 0 to 9 GPa, as well as the effect of zero-point vibrations of phonon modes on the local magnetic moment fluctuations and band structures have been investigated.


\section{COMPUTATIONAL DETAILS}

The first-principles electronic structure calculations were carried out with the Vienna \textit{ab-initio} simulation package,\cite{ab1,ab2} which makes use of the projector augmented wave (PAW) method.\cite{paw} The exchange-correlation functionals were represented by the generalized gradient approximation (GGA) of Perdew-Burke-Ernzerhof (PBE) type.\cite{pbe} In order to describe the vdW interactions not included in the conventional density functional, our calculations adopted the DFT-D2 method\cite{Wu,Grimme} with semi-empirical dispersion energy adding to the Kohn-Sham DFT energy. The energy cutoff for the plane waves was set to 350 eV. The 1 $\times$ 1 $\times$ 1 tetragonal cell of FeSe was used and the integration over the Brillouin zone was performed with a 12 $\times$ 12 $\times$ 12 K-point mesh. The Fermi-level was broadened by the Gaussian smearing method with a width of 0.05 eV. Both the cell parameters and the internal atomic positions were allowed to relax. The system under hydrostatic pressures in a range of 0$-$9 GPa was simulated by assigning the converged trace of the stress tensor to a targeting pressure and minimizing the enthalpy of the system. In the studied pressure range, the structure of the system has not changed to hexagonal.\cite{gpa1} The atoms were allowed to relax until the forces were smaller than 0.01 eV/\AA. After the equilibrium structures were obtained, the frequencies and displacement patterns of the phonon modes were calculated using the dynamical matrix method.\cite{kai} The atomic displacements due to the zero-point vibrations of phonon modes were obtained according to the method of Ref. \onlinecite{zeropoint}. In the present case, the atoms were displaced to the vibrational state with a potential energy of $\hbar \omega_s /$2 for a specified phonon mode $s$, while its normal-mode coordinates can reach two maxima along two opposite directions.

There is another issue in the calculations need to be noted, \textit{i.e.} the magnetic order that we choose in this study. Unlike other iron pnictides such as LaOFeAs\cite{LaOFFeAs} and BaFe$_{2}$As$_{2}$,\cite{BaFeAs} no any long-range magnetic order has been found for $\beta$-FeSe in experiment. However, the strong AFM spin fluctuations were observed for the undoped FeSe. \cite{nmr} Theoretically it is very difficult to directly simulate such a paramagnetic phase by using DFT calculations. Considering that the checkerboard-AFM N\'eel order and the paramagnetic phase share the following important features: (1) local moments around Fe atoms, (2) zero net magnetic moments in a unit cell, and (3) the same space symmetry, the checkerboard-AFM N\'eel state can be thus feasible to properly model the paramagnetic phase in many aspects. \cite{JiWei,Bazhirov} Especially, for the recently grown FeSe monolayer on SrTiO$_3$ with signatures of $T_c$ above 50 K by transport measurement,\cite{xue} the observed shape of Fermi surface in angle-resolved photoemission spectroscopy (ARPES) experiments\cite{zhou,feng} can be reproduced in the DFT calculations by the AFM N\'eel order of FeSe monolayer either without substrate\cite{surface1} or on SrTiO$_3$ substrate.\cite{surface2} In Table I, our calculated lattice parameters for bulk FeSe in nonmagnetic and AFM N\'eel states with or without vdW interactions are compared with the experimental results. As listed in the table, the calculated structure parameters in the AFM N\'eel order with vdW interaction show the best overall agreement with the experimental measurement \cite{experiment1,experiment2} and also yield the better results than the previous calculations by using PBE and hybrid functionals.\cite{kliu} More importantly, the lattice constant along the stacking direction which would change most under pressure is well reproduced. So we choose the AFM N\'eel order to simulate the paramagnetic phase of FeSe in the following studies.


\begin{table}[!b]
\caption{The calculated fully-relaxed lattice parameters for bulk FeSe in nonmagnetic and AFM N\'eel states with or without vdW interactions are listed along with the experimental results.}
\begin{center}
\begin{tabular*}{8.0cm}{@{\extracolsep{\fill}} cccccccc}
\hline
\hline
$ $ & $a$(\AA) & $b$(\AA) & $c$(\AA) & $Z_{Se}$(\AA) \\
\hline
NM \\
No-vdW & 3.679 & 3.679 & 5.999 & 1.385 \\
vdW    & 3.632 & 3.632 & 5.382 & 1.397 \\
\\
N\'eel \\
No-vdW & 3.710 & 3.710 & 6.305 & 1.437 \\
vdW    & 3.654 & 3.654 & 5.471  & 1.436 \\
\hline
Expt.(7K)\cite{experiment1} & 3.765 & 3.754 & 5.479 & 1.462 \\
Expt.(298K)\cite{experiment2} & 3.773 & 3.773 & 5.526 & 1.476  \\
\hline \hline
\end{tabular*}
\end{center}
\end{table}

Regarding the vdW correction to the conventional DFT functionals, the more accurate vdW-optB86b functional,\cite{Klim2} which includes the nonlocal vdW interaction in the exchange and correlation functionals, was also adopted in the studies of FeSe at ambient pressure in order to examine the influence of different vdW approaches. Consistent results of the vdW-optB86b functional and the DFT-D2 method were obtained for the lattice parameters and the interlayer bonding energy ($\sim$26 meV/\AA$^2$). These are in accordance with the previous calculations.\cite{Ricci, Nakamura, Bjorkman}

\section{RESULTS AND ANALYSIS}

\begin{figure}
\includegraphics[angle=0,scale=0.5]{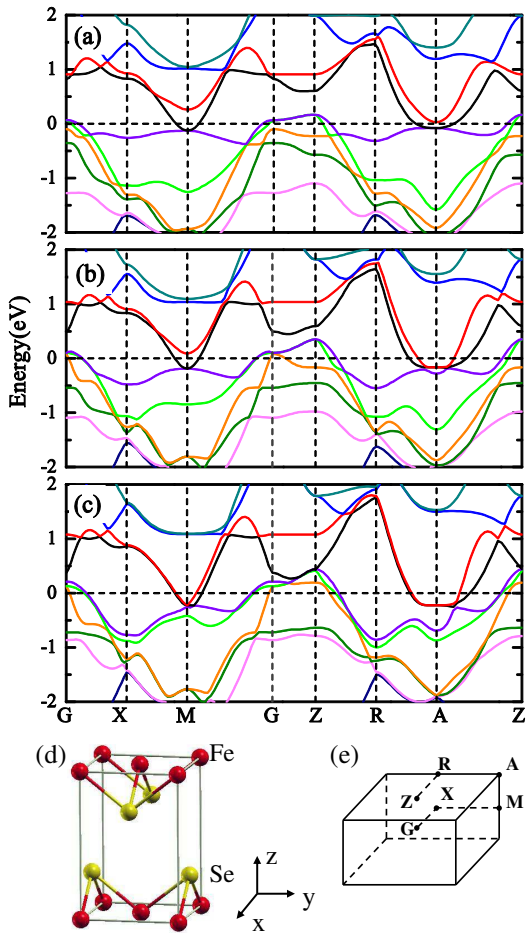}
\caption{(Color online) Electronic band structures of FeSe in AFM N\'eel state under (a)0, (b)5, and (c)6 GPa pressures. The Fermi energy is set to zero. Panels (d) and (e) are the tetragonal cell and the corresponding Brillouin zone, respectively.
}\label{fig1}
\end{figure}

Applying pressure on bulk FeSe results in dramatic changes in its electronic band structures. The energy band near the Fermi level at $\Gamma$ point (labeled by orange line) is occupied at zero pressure [Fig. 1(a)] and it becomes unoccupied at 5 and 6 GPa [Fig. 1(b)(c)]. Based on the analysis of band-decomposed charge density, it is confirmed that this band consists of the $d_{x^2-y^2}$ orbital of Fe at 0 and 5 GPa and changes into the $d_{xz}$/$d_{yz}$ orbital at 6 GPa. Around $M$ point, the unoccupied energy band near Fermi level (in red color) at 0 and 5 GPa becomes occupied at 6 GPa. After the same analysis on charge density, it is ascertained that this band originates from the $d_{x^2-y^2}$ orbital of Fe at zero pressure and becomes the $d_{xz}$/$d_{yz}$ orbital at 5 and 6 GPa. The hole pockets around $\Gamma$ point and the electron pockets around $M$ point are consistent with previous ARPES experiment.\cite{Tamai} In addition to the $\Gamma$ and $M$ points, our calculated band structures also show some obvious changes around the $Z$ and $A$ points. Around $Z$ point, the occupied energy band (labeled by orange line) consisting of Fe $d_{z^2}$ orbital at low pressures becomes unoccupied one with feature of Fe $d_{x^2-y^2}$ orbital at 6 GPa. The changes around $A$ point are similar as that around $M$ point. In one word, the occupations of energy bands around the Fermi level which originate from the $d_{xz}$/$d_{yz}$ and $d_{x^2-y^2}$ orbitals of Fe are very sensitive to pressure. From the analysis on density of states (DOS), contribution from Se atom around the Fermi level is minor. In a recent study combined ARPES experiment and DFT calculations on FeTe$_{0.66}$Se$_{0.34}$,\cite{chen} the contributions from various Fe 3$d$ and Te/Se $p$ orbitals to the bands around Fermi level maily come from the $d_{xz}$/$d_{yz}$ and $d_{x^2-y^2}$ orbitals. In our calculations, the applied pressure makes the crystal lattice constants of FeSe decrease, especially resulting in the collapse of separation between FeSe layers from 5.47 \AA~at 0 GPa to 4.99 \AA~at 6 GPa. This would lead to corresponding changes in electronic properties such as the band structures and orbital occupations.

\begin{figure}
\includegraphics[angle=0,scale=0.32]{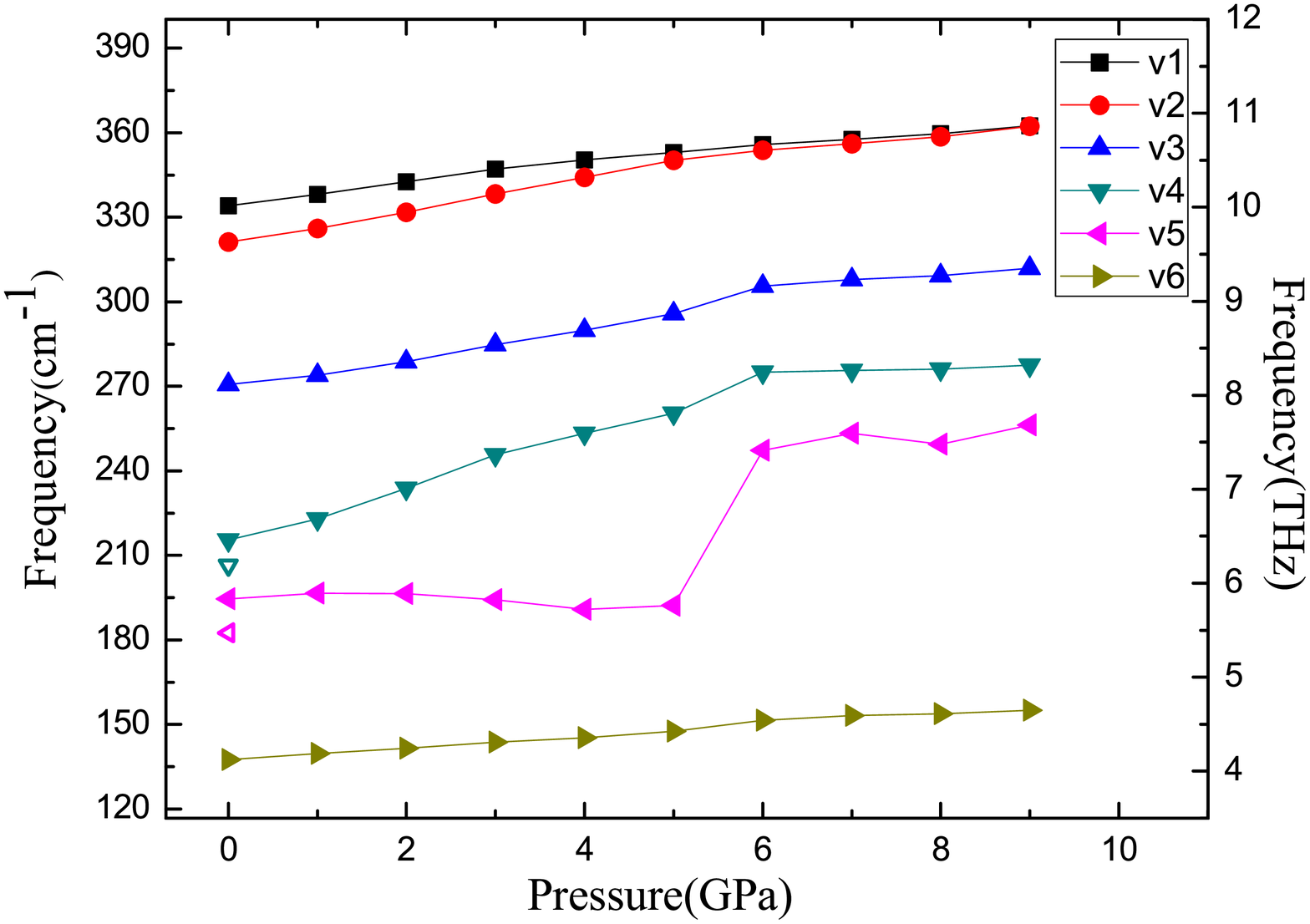}
\caption{(Color online) Calculated phonon frequencies at Brillouin zone center for FeSe in AFM N\'eel order under pressures from 0 to 9 GPa. Hollow triangles at 0 GPa label the corresponding experimental data from Raman scattering measurements.\cite{phonon}}\label{fig2}
\end{figure}

\begin{figure}
\includegraphics[angle=0,scale=0.45]{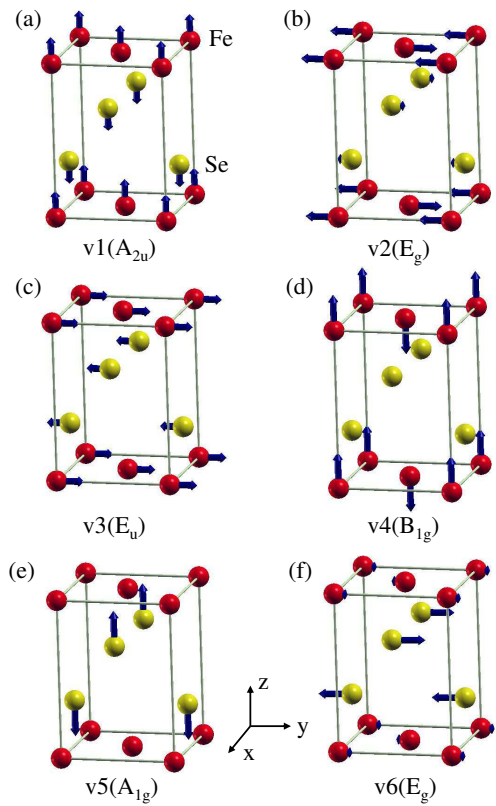}
\caption{(Color online) Atomic displacement patterns for modes (a)v1 ($A_{2u}$), (b)v2 ($E_g$), (c)v3 ($E_u$), (d)v4 ($B_{1g}$), (e)v5 ($A_{1g}$), and (f)v6 ($E_g$) at 6 GPa pressure with corresponding symmetries in the parentheses.}\label{fig3}
\end{figure}

Not only the electronic band structures show large variations with pressure, but also the lattice dynamics demonstrate obvious changes. As shown in Fig. \ref{fig2}, all of the optical phonon frequencies of FeSe at Brillouin zone center increase with pressure. They are labeled in the sequence of energy. The modes v2, v3, and v6 are all doubly degenerate in-plane vibrations. Among all optical phonons, the frequency of mode v5 has a clear sharp jump from 5 to 6 GPa. The hollow triangles in the figure are experimental results of Raman scattering at zero pressure and temperature 7 K,\cite{phonon} which are in the same color as the corresponding calculated values. By comparison between experimental and calculated results, there is about 5\% deviation for the frequencies of modes v4 and v5, which ascertains our theoretical approach using the AFM N\'eel state. The frequencies calculated using fully relaxed structure in nonmagnetic state with vdW interaction show much worse agreement ($\sim$30\% deviation) with experiment. In a recent Raman-scattering measurement on Fe-based superconductor K$_{0.8}$Fe$_{1.6}$Se$_2$ under atmospheric pressure, an anomalous hardening was observed for the A$_g$ mode at $T_c$, indicating a particular type of connection between phonons and superconductivity.\cite{zhanganmin} In our calculations, both the frequency (194.5 cm$^{-1}$) and the symmetry ($A_{1g}$, shown below) of mode v5 for FeSe are similar to the 180 cm$^{-1}$ $A_g$ mode in K$_{0.8}$Fe$_{1.6}$Se$_2$.\cite{zhanganmin} Some experiments on FeSe have shown that depending on the specific methods to apply pressure, a maximum $T_c$ of 37 K can be reached under pressures approximately 6$-$9 GPa.\cite{gpa1,gpa2} The anomalous freuqnecy jump of the phonon mode v5 around the similar pressure range indicates some relationship between this mode and the superconductivity of FeSe.

In order to identify the characteristics of each phonon mode, we plot the atomic displacement patterns at 6 GPa in Fig. \ref{fig3}. The displacement arrows are in the same scale among all panels. For the phonon mode v1, the atoms show the same displacement directions for the same atomic species and the opposite movements for different species. It is an Infrared active $A_{2u}$ mode. Modes v2, v3 and v6 are all doubly degenerate in-plane vibrations, while modes v2 and v6 are Raman active and mode v3 is Infrared active. Both modes v4 and v5 are out-of-plane vibrations. The mode v4 consists of the opposite vertical motions of Fe atoms in the same plane, while the mode v5 involves the coherent motions of Se atoms relative to their adjacent Fe-Fe planes. The atomic displacement patterns are consistent with previous studies.\cite{xia} In Ref. \onlinecite{gpa3}, the authors found that the anion height relative to the Fe-Fe plane is a key factor influencing the $T_c$ of iron-based superconductors. Among all phonon modes of FeSe, atomic displacements in modes v1 and v5 change the Se height mostly. However, in mode v1, the Se atoms above and below the Fe-Fe plane show opposite height changes relative to the Fe-Fe plane and break the original spacial symmetry of FeSe. Meanwhile, the displacements of Se atoms in the $A_{1g}$ mode v5 control precisely the Se height from Fe-Fe plane and keep the same symmetry as that before moving. In a recent time- and angle-resolved photoemission experiments on EuFe$_2$As$_2$,\cite{Rettig} the modulations of electron and hole dynamics due to the $A_{1g}$ phonon was observed. Using femtosecond optical pulses, Kim \textit{et. al.} have also detected the transient magnetic ordering in BaFe$_2$As$_2$ quasi-adiabatically follows the lattice vibrations of $A_{1g}$ mode with a frequency of 5.5 THz.\cite{kim} Due to the sharp increase of phonon frequency for the $A_{1g}$ mode v5 of FeSe at 6 GPa (Fig. 2), it is very tempting to deduce that the $A_{1g}$ phonon mode plays an important role in the $T_c$ increase of FeSe with pressure.

\begin{figure}
\includegraphics[angle=0,scale=0.35]{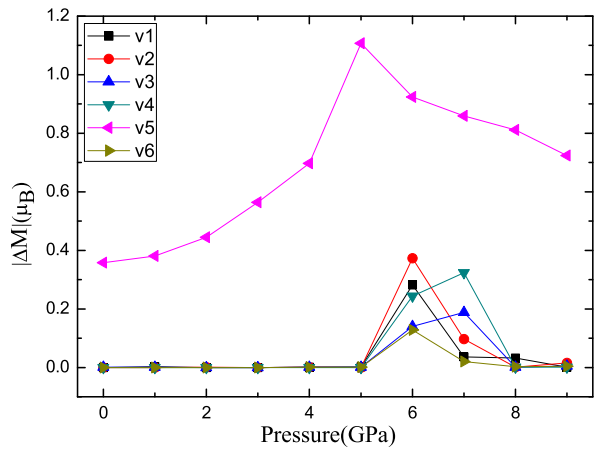}
\caption{(Color online) Changes of local magnetic moment fluctuations on Fe due to zero-point vibrations of different optical phonon modes versus pressure.}\label{fig4}
\end{figure}

Due to the sharp frequency increase of phonon mode v5 from 5 to 6 GPa, zero-point vibrations of optical phonons were simulated in order to study their influence on the spin in FeSe. The FeSe system was set to the vibrational state with zero-point energy of $\hbar$w$_s$/2 in a specified phonon mode $s$, while the normal-mode coordinate could reach maxima along two opposite directions. So there are two displacement patterns for each mode at each pressure. We plot the difference of local magnetic moment on Fe between the two displacement patterns of each mode at various pressures (Fig. 4), namely $|\bigtriangleup M|$ = $|M_+ - M_-|$, with $M_+$ being the local magnetic moment on Fe for one displacement and $M_-$ the other. From Fig. \ref{fig4}, $|\bigtriangleup M|$ caused by the zero-point vibration of mode v5 has much bigger values than all other phonon modes from 0 to 9 GPa and it reaches maximum value at 5 GPa. For zero-point vibrations induced by other phonon modes, $|\bigtriangleup M|$ are not zero only around 6 GPa. So it can be concluded that all phonon modes enhance the AFM fluctuations around 6 GPa. This is in accordance with the experimental observations that application of pressure on FeSe enhances AFM spin fluctuations.\cite{nmr} As the pressure increase, the occupations of energy bands around Fermi level consisting of Fe $d$ orbitals demonstrate large variations (Fig. 1). The sensitive occupations of these Fe $d$ orbitals to lattice vibrations are responsible for the local magnetic moment fluctuations. In addition, the local magnetic moment $M$ on Fe in FeSe at equilibrium structure gets smaller and smaller with increasing pressure, which can be learned that the pressure suppresses the AFM state. This also agrees with the viewpoint that superconductivity is induced by doping charge carriers into the parent compound to suppress the AFM state in both the cuprate\cite{phys-b,p-a-lee} and the iron-based\cite{sc1,sc2,sc3,sc4,rmp1,rmp2} superconductors.  Because of the anomalous frequency increase of phonon mode v5 (Fig. \ref{fig2}) and the enhanced spin fluctuations around 5$-$6 GPa (Fig. \ref{fig4}), it is arrived that spin-phonon coupling exists and plays an important role in FeSe under pressure.

\begin{figure}
\includegraphics[angle=0,scale=0.38]{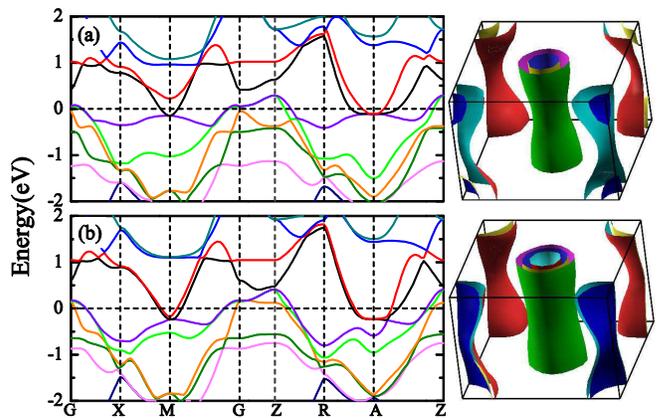}
\caption{(Color online) Electronic band structures and Fermi surfaces of FeSe in AFM N\'eel state as a result of zero-point vibration of the $A_{1g}$ mode at 6 GPa. The Fermi energy is set to zero. Panels (a) and (b) correspond to movements of atom Se far away from and close to the Fe-Fe plane, respectively. The high symmetry points in Brillouin zone are the same as that in Fig. \ref{fig1}(e).}\label{fig5}
\end{figure}

Although the atomic displacements due to the zero-point vibration of the $A_{1g}$ mode are very small (about 0.03 \AA), the effect on the electronic band structure is obvious. Fig. \ref{fig5} is the corresponding band structures when Se moves (a) away from and (b) close to the Fe-Fe plane at 6 GPa. At $\Gamma$ point, the band around Fermi level labeled by orange line is occupied when Se moves away from the Fe-Fe plane [Fig. 5(a)] and becomes unoccupied when Se moves close [Fig. 5(b)]. Through analysis from the band-decomposed charge density, this band consists of the Fe $d_{x^2-y^2}$ orbital in Fig. 5(a) and changes into the $d_{xz}$/$d_{yz}$ orbital in Fig. 5(b). Around $M$ point, the energy band represented by red line shows opposite change to that at $\Gamma$ point. After the same analysis, it is confirmed that the band around $M$ point comes from the $d_{x^2-y^2}$ and $d_{xz}$/$d_{yz}$ orbitals of Fe in Fig. 5(a) and Fig. 5(b), respectively. However, the band change around the $A$ point is a little different from $M$ point. For $A$ point, the energy band (in red color) shifts down with decreased energy when Se moves close to the Fe-Fe plane. The three-dimensional views can be seen more clearly from the shape changes of Fermi surfaces as plotted adjacent to the corresponding band structures. Small changes of crystal structure due to zero-point vibration alone could not affect the energy band so much unless some other physical mechanism is included in this process. Nonmagnetic calculations have been performed to ascertain our speculations. Only in spin-polarized calculations, the zero-point vibrations have great impact on the energy bands. The dramatic changes of band structures and Fermi surfaces due to the zero-point vibrations of $A_{1g}$ mode in AFM N\'eel state at 6 GPa further support that spin-phonon coupling plays a key role in FeSe under pressure.

\section{DISCUSSION and Summary}

Our above calculations show that the zero-point vibration of the $A_{1g}$ mode of FeSe, which relates with the Se height from the Fe-Fe plane, induces large fluctuations of local magnetic moment on Fe and the spin fluctuations caused by spin-phonon coupling are further enhanced under pressure. In recent experiments by means of magnetization and neutron power diffraction, a clear isotope effect on $T_c$ is observed for bulk FeSe, which highlights the role of the lattice in the paring mechanism.\cite{Khasanov} In a Raman-scattering measurement on K$_{0.8}$Fe$_{1.6}$Se$_2$, an anomaly at $T_c$ in the 180 cm$^{-1}$ $A_g$ mode is observed, which indicates a rather specific type of electron-phonon coupling.\cite{zhanganmin} For the recently grown monolayer FeSe on SrTiO$_3$ substrate showing signatures of $T_c$ above 50 K,\cite{xue} the screening due to the SrTiO$_3$ ferroelectric phonons on Cooper paring in monolayer FeSe is proposed to significantly enhance the energy scale of Cooper paring and even change the paring symmetry.\cite{xiang} From first-principles studies on FeSe and KFe$_2$Se$_2$, the estimates of $T_c$ based on spin-resolved coupling values show around a twofold increase than that from non-spin-resolved configurations.\cite{Bazhirov} These experimental and theoretical studies all suggest that the effect of phonon could not be completely ignored in the paring mechanism of FeSe.

Not only in iron chalcogenides, there are also evidences of phonon effects on the unconventional superconductivity in iron pnictides and cuprates. On the experimental side, for SmFeAsO$_{1-x}$F$_x$ and Ba$_{1-x}$K$_x$Fe$_2$As$_2$ systems, the iron isotope substitution shows the same effect on $T_c$ and the spin-density wave transition temperature $T_{SDW}$, suggesting that strong magnon-phonon coupling exists.\cite{liurh} Using ultrashort and intense optical pulses probe, ultrafast transient spin-density-wave order develops in the normal state of BaFe$_2$As$_2$ and is driven by coherent lattice vibrations even without breaking the crystal symmetry, which attests a pronounced spin-phonon coupling in pnictides.\cite{kim} From ARPES probe of the electron dynamics in three different families of copper oxide superconductors,\cite{shen} which share a common thread of spin-fluctuation mediated pairing as iron-based superconductors,\cite{rmp2} it is found that an abrupt change of electron velocity at 50-80 meV can not be explained by any known process other than the coupling to phonon is included.\cite{shen} On the theoretical side, Yildirim finds strong coupling of the on-site Fe-magnetic moment with the As-As bonding in iron-pnictide superconductors from first-principles calculations.\cite{Yildirim} For computational studies on doped LaFeAsO, the coupling magnetism with vibrations is also found to induce anharmonicities and an electron-phonon interaction much larger than in the paramagnetic state.\cite{Yndurain1,Yndurain2} From these studies, the spin-phonon coupling is evidently ubiquitous in iron-based superconductors.

The spin-fluctuation mediated pairing is common in iron-based superconductors and other unconventional superconducting materials.\cite{rmp1,rmp2} Regarding the nature of the magnetism in iron-based superconductors, there are basically two contradictive views. The one is based on itinerant electron picture,\cite{mazin} in which the Fermi surface nesting is responsible for the AFM order. On the contrary, the other one is based on local moment interactions which can be described by the $J_1$-$J_2$ frustrated Heisenberg model.\cite{yildirim,si,ma1} And it was further shown~\cite{ma1} that the underlying driving force herein is the anion-bridged AFM superexchange interaction between a pair of the next-nearest-neighboring fluctuating Fe local moments embedded in itinerant electrons. There are now more and more evidences in favor of the fluctuating Fe local moment picture. Especially, the inelastic neutron scattering experiments have shown that the low-energy magnetic excitations can be well described by the spin waves based on the $J_1$-$J_2$ Heisenberg model.\cite{dai-0,dai-1} Our calculated results of FeSe here show that the fluctuations of local magnetic moment induced by the zero-point vibration of $A_{1g}$ phonon mode are significant and are further enhanced under pressure. Eventhough the direct electron-phonon coupling calculations both without\cite{Subedi} and with\cite{Bazhirov} spin polarization effects cannot account for the experimentally observed $T_{c}$ of FeSe, the phonon could play an indirect role through spin-phonon coupling. Our results suggest the effect of phonon should be included when unraveling the paring mechanism in iron-based superconductors.

To summarize, the variation of band structures and the phonon frequencies of FeSe under 0 to 9 GPa hydrostatic pressure, as well as the effect of the zero-point vibrations on the local magnetic moment fluctuations and band structures, have been investigated by using DFT calculations with vdW corrections. With applied pressure, the energy bands consisting of Fe $d$ orbitals around the Fermi level show obvious shifts and occupation changes. At the same time, the frequencies of all optical phonon modes at Brillouin zone center increase with pressure. Among these phonon modes, the $A_{1g}$ mode related to the Se height from Fe-Fe plane shows a clear frequency jump from 5 to 6 GPa. This is around the similar pressure range within which the highest $T_c$ is observed for FeSe in experiment. Compared with other phonon modes, the zero-point vibration of the $A_{1g}$ mode also induces the strongest fluctuation of local magnetic moment on Fe from 0 to 9 GPa and the fluctuation reaches maximum at 5 GPa. The enhanced fluctuations of local magnetic moment may be favorable to promote the $T_c$. These results highlight the role of spin-phonon coupling when exploring the superconducting mechanism of iron-based superconductors.


\begin{acknowledgments}

We wish to thank Professor Shiwu Gao for helpful communications. This work is supported by National Natural Science Foundation of China (Grant Nos. 11004243, 11190024, and 51271197) and National Program for Basic Research of MOST of China (Grant No. 2011CBA00112). Computational resources have been provided by the Physical Laboratory of High Performance Computing at Renmin University of China. The atomic structures and Fermi surfaces were prepared with the XCRYSDEN program.\cite{kokalj}

\end{acknowledgments}

\end{document}